\documentclass[conference]{IEEEtran}
\IEEEoverridecommandlockouts

\usepackage{cite}
\usepackage{amsmath,amssymb,amsfonts}
\usepackage{bm}
\usepackage{graphicx}
\usepackage{textcomp}
\usepackage{xcolor}
\usepackage{tikz}
\usepackage{pgfplots}
\pgfplotsset{compat=1.18}
\usepackage{etoolbox}
\usepackage{algorithm}
\usepackage{algorithmicx}
\usepackage{algpseudocode}
\usepackage[T1]{fontenc}
\usepackage[utf8]{inputenc}
\usepackage[nolist,printonlyused]{acronym}
\usepackage{tabularx}
\usepackage{array}
\usepackage[labelfont=bf,font=small]{caption}
\usepackage{soul}
\usepackage{xcolor}

\usetikzlibrary{arrows.meta,calc,positioning}

\makeatletter
\patchcmd{\maketitle}{\vskip 1.5em}{\vskip 0.8em}{}{}
\makeatother
\IEEEaftertitletext{\vspace{-1.0\baselineskip}}

\def\BibTeX{{\rm B\kern-.05em{\sc i\kern-.025em b}\kern-.08em
    T\kern-.1667em\lower.7ex\hbox{E}\kern-.125emX}}

\author{}
\date{}

\begin{acronym}
\acro{NF}{near-field}
    \acro{PR}{partial relaxation}
\acro{PR-ML}{partial-relaxation maximum likelihood}
\acro{PR-CF}{partial-relaxation covariance fitting}
  \acro{ULA}{uniform linear array}
  \acro{CRB}{Cram\'er--Rao bound}
  \acro{CCRB}{constrained Cram\'er--Rao bound}
  \acro{ML}{maximum likelihood}
  \acro{SNR}{signal-to-noise ratio}
\acro{2D-MUSIC}{two-dimensional MUltiple SIgnal Classification}
  \acro{PR-CRB}{partial-relaxation Cram\'er--Rao bound}
  \acro{PR-CCRB}{partial-relaxation constrained Cram\'er--Rao bound}
\acro{FIM}{Fisher information matrix}
\acro{PRCRBp}[PR-CRB\ensuremath{_{\mathrm{p}}}]
{Partial-Relaxation Cram\'er--Rao Bound for Known Pilots}

\acro{PRCRBu}[PR-CRB\ensuremath{_{\mathrm{u}}}]
{Partial-Relaxation Cram\'er--Rao Bound for Unknown Symbols}
\acro{NMS}[NMS]{non-maximum suppression}

  \acro{2G}{Second Generation}
  \acro{3G}{3$^\text{rd}$~Generation}
  \acro{3GPP}{3$^\text{rd}$~Generation Partnership Project}
  \acro{4G}{4$^\text{th}$~Generation}
  \acro{5G}{5$^\text{th}$~Generation}
  \acro{AA}{Antenna Array}
  \acro{AC}{Admission Control}
  \acro{AD}{Attack-Decay}
  \acro{ADSL}{Asymmetric Digital Subscriber Line}
	\acro{AHW}{Alternate Hop-and-Wait}
  \acro{AMC}{Adaptive Modulation and Coding}
  \acro{AoA}{angle of arrival}
  \acro{AoD}{angle of departure}
	\acro{AP}{Access Point}
  \acro{APA}{Adaptive Power Allocation}
  \acro{AR}{autoregressive}
  \acro{ARMA}{Autoregressive Moving Average}
  \acro{ATES}{Adaptive Throughput-based Efficiency-Satisfaction Trade-Off}
  \acro{AWGN}{additive white Gaussian noise}
  \acro{BB}{Branch and Bound}
  \acro{BD}{Block Diagonalization}
  \acro{BER}{bit error rate}
  \acro{BF}{Best Fit}
  \acro{BLER}{BLock Error Rate}
  \acro{BPC}{Binary power control}
  \acro{BPSK}{Binary Phase-Shift Keying}
  \acro{BPA}{Best \ac{PDPR} Algorithm}
  \acro{BRA}{Balanced Random Allocation}
  \acro{BCRB}{Bayesian Cram\'{e}r-Rao Bound}
  \acro{BS}{base station}
  \acro{CAP}{Combinatorial Allocation Problem}
  \acro{CAPEX}{Capital Expenditure}
  \acro{CBF}{Coordinated Beamforming}
  \acro{CBR}{Constant Bit Rate}
  \acro{CBS}{Class Based Scheduling}
  \acro{CC}{Congestion Control}
  \acro{CDF}{Cumulative Distribution Function}
  \acro{CDMA}{Code-Division Multiple Access}
  \acro{CL}{Closed Loop}
  \acro{CLPC}{Closed Loop Power Control}
  \acro{CNR}{Channel-to-Noise Ratio}
  \acro{CPA}{Cellular Protection Algorithm}
  \acro{CPICH}{Common Pilot Channel}
  \acro{CoMP}{Coordinated Multi-Point}
  \acro{CQI}{Channel Quality Indicator}
  \acro{CRM}{Constrained Rate Maximization}
	\acro{CRN}{Cognitive Radio Network}
  \acro{CS}{Coordinated Scheduling}
  \acro{CSI}{channel state information}
  \acro{CSIR}{channel state information at the receiver}
  \acro{CSIT}{channel state information at the transmitter}
  \acro{CUE}{cellular user equipment}
  \acro{D2D}{device-to-device}
  \acro{DCA}{Dynamic Channel Allocation}
  \acro{DE}{Differential Evolution}
  \acro{DFT}{Discrete Fourier Transform}
  \acro{DIST}{Distance}
  \acro{DL}{downlink}
  \acro{DMA}{Double Moving Average}
	\acro{DMRS}{demodulation reference signal}
  \acro{D2DM}{D2D Mode}
  \acro{DMS}{D2D Mode Selection}
  \acro{DPC}{Dirty Paper Coding}
  \acro{DRA}{Dynamic Resource Assignment}
  \acro{DSA}{Dynamic Spectrum Access}
  \acro{DSM}{Delay-based Satisfaction Maximization}
  \acro{ECC}{Electronic Communications Committee}
  \acro{EFLC}{Error Feedback Based Load Control}
  \acro{EI}{Efficiency Indicator}
  \acro{eNB}{Evolved Node B}
  \acro{EPA}{Equal Power Allocation}
  \acro{EPC}{Evolved Packet Core}
  \acro{EPS}{Evolved Packet System}
  \acro{ESPRIT}{estimation of signal parameters via rotational invariance}
  \acro{E-UTRAN}{Evolved Universal Terrestrial Radio Access Network}
  \acro{ES}{Exhaustive Search}
  \acro{FDD}{frequency division duplexing}
  \acro{FDM}{Frequency Division Multiplexing}
  \acro{FER}{Frame Erasure Rate}
  \acro{FF}{Fast Fading}
  \acro{FI}{Fisher information}
  \acro{FIM}{Fisher information matrix}
  \acro{FSB}{Fixed Switched Beamforming}
  \acro{FST}{Fixed SNR Target}
  \acro{FTP}{File Transfer Protocol}
  \acro{GA}{Genetic Algorithm}
  \acro{GBR}{Guaranteed Bit Rate}
  \acro{GLR}{Gain to Leakage Ratio}
  \acro{GOS}{Generated Orthogonal Sequence}
  \acro{GPL}{GNU General Public License}
  \acro{GRP}{Grouping}
  \acro{HARQ}{Hybrid Automatic Repeat Request}
  \acro{HMS}{Harmonic Mode Selection}
  \acro{HOL}{Head Of Line}
  \acro{HSDPA}{High-Speed Downlink Packet Access}
  \acro{HSPA}{High Speed Packet Access}
  \acro{HTTP}{HyperText Transfer Protocol}
  \acro{ICMP}{Internet Control Message Protocol}
  \acro{ICI}{Intercell Interference}
  \acro{ID}{Identification}
  \acro{ISAC}{integrated sensing and communication}
  \acro{IETF}{Internet Engineering Task Force}
  \acro{ILP}{Integer Linear Program}
  \acro{JRAPAP}{Joint RB Assignment and Power Allocation Problem}
  \acro{UID}{Unique Identification}
  \acro{IID}{Independent and Identically Distributed}
  \acro{IIR}{Infinite Impulse Response}
  \acro{ILP}{Integer Linear Problem}
  \acro{IMT}{International Mobile Telecommunications}
  \acro{INV}{Inverted Norm-based Grouping}
	\acro{IoT}{Internet of Things}
  \acro{IP}{Internet Protocol}
  \acro{IPv6}{Internet Protocol Version 6}
  \acro{ISD}{Inter-Site Distance}
  \acro{ISI}{Inter Symbol Interference}
  \acro{ITU}{International Telecommunication Union}
  \acro{JOAS}{Joint Opportunistic Assignment and Scheduling}
  \acro{JOS}{Joint Opportunistic Scheduling}
  \acro{JP}{Joint Processing}
	\acro{JS}{Jump-Stay}
  \acro{KKT}{Karush-Kuhn-Tucker}
  \acro{KPI}{key performance indicator}
  \acro{L3}{Layer-3}
  \acro{LAC}{Link Admission Control}
  \acro{LA}{Link Adaptation}
  \acro{LC}{Load Control}
  \acro{LOS}{Line of Sight}
  \acro{LP}{Linear Programming}
  \acro{LS}{least squares}
  \acro{LTE}{Long Term Evolution}
  \acro{LTE-A}{LTE-Advanced}
  \acro{LTE-Advanced}{Long Term Evolution Advanced}
  \acro{M2M}{Machine-to-Machine}
  \acro{MAC}{Medium Access Control}
  \acro{MANET}{Mobile Ad hoc Network}
  \acro{MC}{Modular Clock}
  \acro{MCS}{Modulation and Coding Scheme}
  \acro{MDB}{Measured Delay Based}
  \acro{MDI}{Minimum D2D Interference}
  \acro{MF}{Matched Filter}
  \acro{MG}{Maximum Gain}
  \acro{MH}{Multi-Hop}
  \acro{MIMO}{multiple input multiple output}
  \acro{MINLP}{Mixed Integer Nonlinear Programming}
  \acro{MIP}{Mixed Integer Programming}
  \acro{MISO}{Multiple Input Single Output}
  \acro{ML}{maximum likelihood}
  \acro{MLE}{maximum likelihood estimator}
  \acro{MLWDF}{Modified Largest Weighted Delay First}
  \acro{MME}{Mobility Management Entity}
  \acro{MMSE}{minimum mean squared error}
  \acro{MOS}{Mean Opinion Score}
  \acro{MPF}{Multicarrier Proportional Fair}
  \acro{MRA}{Maximum Rate Allocation}
  \acro{MR}{Maximum Rate}
  \acro{MRC}{Maximum Ratio Combining}
  \acro{MRT}{Maximum Ratio Transmission}
  \acro{MRUS}{Maximum Rate with User Satisfaction}
  \acro{MS}{mobile station}
  \acro{MSE}{mean squared error}
  \acro{MSI}{Multi-Stream Interference}
  \acro{MTC}{Machine-Type Communication}
  \acro{MTSI}{Multimedia Telephony Services over IMS}
  \acro{MTSM}{Modified Throughput-based Satisfaction Maximization}
  \acro{MU-MIMO}{multiuser multiple input multiple output}
  \acro{MU}{multi-user}
  \acro{MUSIC}{MUltiple SIgnal Classification}
  \acro{NAS}{Non-Access Stratum}
  \acro{NB}{Node B}
  \acro{NE}{Nash equilibrium}
  \acro{NCL}{Neighbor Cell List}
  \acro{NLP}{Nonlinear Programming}
  \acro{NLOS}{Non-Line of Sight}
  \acro{NMSE}{normalized mean squared error}
  \acro{NORM}{Normalized Projection-based Grouping}
  \acro{NP}{Non-Polynomial Time}
  \acro{NRT}{Non-Real Time}
  \acro{NSPS}{National Security and Public Safety Services}
  \acro{O2I}{Outdoor to Indoor}
  \acro{OFDMA}{orthogonal frequency division multiple access}
  \acro{OFDM}{orthogonal frequency division multiplexing}
  \acro{OFPC}{Open Loop with Fractional Path Loss Compensation}
	\acro{O2I}{Outdoor-to-Indoor}
  \acro{OL}{Open Loop}
  \acro{OLPC}{Open-Loop Power Control}
  \acro{OL-PC}{Open-Loop Power Control}
  \acro{OPEX}{Operational Expenditure}
  \acro{ORB}{Orthogonal Random Beamforming}
  \acro{JO-PF}{Joint Opportunistic Proportional Fair}
  \acro{OSI}{Open Systems Interconnection}
  \acro{PAIR}{D2D Pair Gain-based Grouping}
  \acro{PAPR}{Peak-to-Average Power Ratio}
  \acro{P2P}{Peer-to-Peer}
  \acro{PC}{Power Control}
  \acro{PCI}{Physical Cell ID}
  \acro{PDF}{Probability Density Function}
  \acro{PDPR}{pilot-to-data power ratio}
  \acro{PER}{Packet Error Rate}
  \acro{PF}{Proportional Fair}
  \acro{P-GW}{Packet Data Network Gateway}
  \acro{PL}{Pathloss}
  \acro{PPR}{pilot power ratio}
  \acro{PRB}{physical resource block}
  \acro{PROJ}{Projection-based Grouping}
  \acro{ProSe}{Proximity Services}
  \acro{PS}{Packet Scheduling}
  \acro{PSAM}{pilot symbol assisted modulation}
  \acro{PSK}{phase-shift keying}
  \acro{PSO}{Particle Swarm Optimization}
  \acro{PZF}{Projected Zero-Forcing}
  \acro{QAM}{Quadrature Amplitude Modulation}
  \acro{QoS}{Quality of Service}
  \acro{QPSK}{Quadri-Phase Shift Keying}
  \acro{RAISES}{Reallocation-based Assignment for Improved Spectral Efficiency and Satisfaction}
  \acro{RAN}{Radio Access Network}
  \acro{RA}{Resource Allocation}
  \acro{RAT}{Radio Access Technology}
  \acro{RATE}{Rate-based}
  \acro{RB}{resource block}
  \acro{RBG}{Resource Block Group}
  \acro{REF}{Reference Grouping}
  \acro{RLC}{Radio Link Control}
  \acro{RM}{Rate Maximization}
  \acro{RNC}{Radio Network Controller}
  \acro{RND}{Random Grouping}
  \acro{RRA}{Radio Resource Allocation}
  \acro{RRM}{Radio Resource Management}
  \acro{RSCP}{Received Signal Code Power}
  \acro{RSRP}{Reference Signal Receive Power}
  \acro{RSRQ}{Reference Signal Receive Quality}
  \acro{RR}{Round Robin}
  \acro{RRC}{Radio Resource Control}
  \acro{RSSI}{Received Signal Strength Indicator}
  \acro{RT}{Real Time}
  \acro{RU}{Resource Unit}
  \acro{RUNE}{RUdimentary Network Emulator}
  \acro{RV}{Random Variable}
  \acro{Rx}{receiver}
  \acro{SAC}{Session Admission Control}
  \acro{SCM}{Spatial Channel Model}
  \acro{SC-FDMA}{Single Carrier - Frequency Division Multiple Access}
  \acro{SD}{Soft Dropping}
  \acro{S-D}{Source-Destination}
  \acro{SDPC}{Soft Dropping Power Control}
  \acro{SDMA}{Space-Division Multiple Access}
  \acro{SER}{Symbol Error Rate}
  \acro{SES}{Simple Exponential Smoothing}
  \acro{S-GW}{Serving Gateway}
  \acro{SINR}{signal-to-interference-plus-noise ratio}
  \acro{SI}{Satisfaction Indicator}
  \acro{SIP}{Session Initiation Protocol}
  \acro{SISO}{single input single output}
  \acro{SIMO}{Single Input Multiple Output}
  \acro{SIR}{signal-to-interference ratio}
  \acro{SLNR}{Signal-to-Leakage-plus-Noise Ratio}
  \acro{SMA}{Simple Moving Average}
  \acro{SNR}{signal-to-noise ratio}
  \acro{SORA}{Satisfaction Oriented Resource Allocation}
  \acro{SORA-NRT}{Satisfaction-Oriented Resource Allocation for Non-Real Time Services}
  \acro{SORA-RT}{Satisfaction-Oriented Resource Allocation for Real Time Services}
  \acro{SPF}{Single-Carrier Proportional Fair}
  \acro{SRA}{Sequential Removal Algorithm}
  \acro{SRS}{Sounding Reference Signal}
  \acro{SSB}{synchronisation signal block}
  \acro{SU-MIMO}{single-user multiple input multiple output}
  \acro{SU}{Single-User}
  \acro{SVD}{Singular Value Decomposition}
  \acro{TCP}{Transmission Control Protocol}
  \acro{TDD}{time division duplexing}
  \acro{TDMA}{Time Division Multiple Access}
  \acro{TETRA}{Terrestrial Trunked Radio}
  \acro{TP}{Transmit Power}
  \acro{TPC}{Transmit Power Control}
  \acro{TTI}{Transmission Time Interval}
  \acro{TTR}{Time-To-Rendezvous}
  \acro{TSM}{Throughput-based Satisfaction Maximization}
  \acro{TU}{Typical Urban}
  \acro{Tx}{transmitter}
  \acro{UE}{user equipment}
  \acro{UEPS}{Urgency and Efficiency-based Packet Scheduling}
  \acro{UL}{uplink}
  \acro{ULA}{uniform linear array}
  \acro{UMTS}{Universal Mobile Telecommunications System}
  \acro{URI}{Uniform Resource Identifier}
  \acro{URM}{Unconstrained Rate Maximization}
  \acro{UT}{user terminal}
  \acro{VR}{Virtual Resource}
  \acro{VoIP}{Voice over IP}
  \acro{WAN}{Wireless Access Network}
  \acro{WCDMA}{Wideband Code Division Multiple Access}
  \acro{WF}{Water-filling}
  \acro{WiMAX}{Worldwide Interoperability for Microwave Access}
  \acro{WINNER}{Wireless World Initiative New Radio}
  \acro{WLAN}{Wireless Local Area Network}
  \acro{WMPF}{Weighted Multicarrier Proportional Fair}
  \acro{WPF}{Weighted Proportional Fair}
  \acro{WSN}{Wireless Sensor Network}
  \acro{WWW}{World Wide Web}
  \acro{XIXO}{(Single or Multiple) Input (Single or Multiple) Output}
  \acro{ZF}{zero-forcing}
  \acro{ZMCSCG}{Zero Mean Circularly Symmetric Complex Gaussian}
\end{acronym}

\begin{document}

\title{Low-Complexity Multi-User Non-Line-of-Sight Channel Estimation}




\author{
\IEEEauthorblockN{
Mohammad Abu Aqoulah$^{1,4}$,
Do\u{g}a G\"urg\"uno\u{g}lu$^{2}$,
G\'abor Fodor$^{3,4}$,
Gonzalo Seco-Granados$^{1}$
}
\IEEEauthorblockA{$^{1}$Universitat Aut\`onoma de Barcelona (UAB), CERES, Barcelona, Spain}
\IEEEauthorblockA{$^{2}$ASELSAN Inc., Ankara, Türkiye}
\IEEEauthorblockA{$^{3}$School of EECS, KTH Royal Institute of Technology, Stockholm, Sweden}
\IEEEauthorblockA{$^{4}$Ericsson Research, Stockholm, Sweden}
\IEEEauthorblockA{Emails: \{Mohammad.AbuAqoulah@uab.cat, dgurgunoglu@aselsan.com, gabor.fodor@ericsson.com, gonzalo.seco@uab.cat\}}
\thanks{This work has been partially supported by the European Union's Horizon Europe Marie Sklodowska-Curie Innovative Training Networks HORIZON-MSCA-2022-DN-01 call, under Grant Agreement "MiFuture" No.\ 101119643, by the Spanish Agency of Research (AEI) under grant PID2023-152820OB-I00, funded by MICIU/AEI/10.13039/501100011033 and ERDF/EU, and by the Catalan Government under the AGAUR-ICREA Academia Program.
G.\ Fodor was also supported by the Swedish Strategic Research (SSF) grant for the FUS21-0004 SAICOM project.}
}

\maketitle

\vspace{0.75em}
\begin{abstract}
Future radio access networks are expected to rely on large-aperture antenna arrays, for which an increasing portion of the coverage region may fall within the radiative near-field. In this regime, conventional far-field models become inaccurate, and the received spatial signature depends jointly on the propagation range and azimuth, making reliable uplink channel acquisition a key physical-layer challenge. Many existing near-field estimation works rely on simplified line-of-sight-dominant or single-path channel models, which fail to capture practical non-line-of-sight (NLoS) multipath environments. In contrast to prior single-path near-field partial relaxation (PR) formulations, in this paper, we develop a PR-based framework for near-field NLoS uplink channel estimation by modeling each user channel as a superposition of multiple spherical-wave components characterized by their ranges and azimuths. For the known-pilot case, we develop a greedy PR-based maximum likelihood estimator that iteratively extracts dominant propagation paths while mitigating multi-user interference. For the unknown-symbol case, we propose a PR-based rank-adaptive covariance-fitting approach that captures the multipath structure. We further derive the corresponding Cram\'er–Rao bounds for both cases. Numerical results show that the proposed multipath PR-based methods achieve strong estimation performance, remain close to the corresponding bounds, and outperform near-field two-dimensional multiple signal classification across the considered scenarios, supporting their relevance for future large-aperture uplink systems.
\end{abstract}



\section{Introduction}

\Ac{MIMO} systems have become a key enabler of modern wireless networks by exploiting spatial degrees of freedom through multi-antenna transmission and reception. In the evolution toward 6G and IMT-2030, future radio access networks are expected to support more demanding communication, localization, and sensing capabilities, for which large antenna arrays are among the key enabling technologies \cite{Liu2025ITU6G}. In addition to improving spectral efficiency via beamforming and spatial multiplexing, large antenna arrays also enable emerging capabilities such as localization and sensing \cite{Andrews:24,Yang:25,Yazid:25b}. These functionalities, however, rely critically on accurate \ac{CSI}. As a result, designing channel estimation techniques that provide reliable CSI while maintaining reasonable pilot overhead and computational complexity remains a central challenge in contemporary and future wireless systems \cite{Andrews:24}.

As antenna arrays continue to grow in size and aperture, the underlying propagation characteristics of wireless channels change significantly. In particular, users located close to the base station may fall within the radiative \ac{NF}, where the wavefront curvature across the array becomes non-negligible \cite{Zhang2023NearField}. In this regime, the received phase profile depends jointly on both the propagation distance and the angle of arrival, which makes conventional far-field planar-wave models inaccurate. Consequently, channel estimation techniques must explicitly account for this \ac{NF} structure in order to achieve reliable performance, especially in multi-user uplink scenarios where interference and limited snapshot availability further complicate the estimation task.

Subspace-based methods are widely used for parameter estimation in array processing. In \ac{NF} scenarios, the \ac{2D-MUSIC} algorithm has been extended to spherical-wave models to jointly estimate range and angle \cite{gurgunoglu2025_2dmusic}. However, these approaches rely on the estimation of the sample covariance matrix and may suffer from subspace leakage when the number of snapshots is limited or the \ac{SNR} is low. Their performance can further degrade in multi-user uplink scenarios due to strong interference and closely spaced users.

To better handle multi-user interference, the partial relaxation (PR) principle has been proposed for array parameter estimation. In this framework, the desired user is modeled parametrically, while the remaining users are absorbed into a relaxed nuisance component \cite{TrinhHoang2018PRFramework}. Prior studies have shown that PR-based estimators can outperform classical subspace methods \cite{TrinhHoang2018PRFramework,Schenck2019PRCF,jade_pr_paper}. Recently, this framework was extended to radiative \ac{NF} channel estimation under a spherical-wave model \cite{AbuAqoulahICC}. However, that work considered a simplified single-path channel model, leaving practical \ac{NF} NLoS multipath propagation insufficiently explored.

This limitation is significant because, in NLoS near-field channels, the desired-user channel may contain multiple spherical-wave components with different ranges, azimuths, and complex gains. This increases the parameter dimension and introduces path-selection ambiguities that are absent in the single-path formulation. Therefore, a clear gap remains in extending PR-based near-field estimation from the single-path setting to practical NLoS multipath channels.

Motivated by this gap, this paper extends the \ac{PR} framework to \ac{NF} NLoS multipath channel estimation. The desired-user channel is modeled as a superposition of \ac{NF} paths, each parameterized by its range and azimuth under the spherical-wave model. Compared with the single-path PR formulation in \cite{AbuAqoulahICC}, the proposed formulation explicitly accounts for multiple propagation paths and adaptively extracts the dominant path components of the desired user while relaxing the remaining multi-user interference. Based on this model, we develop \ac{PR-ML} for known pilots and \ac{PR-CF} for unknown symbols, and derive the corresponding performance bounds, namely \ac{PRCRBp} and \ac{PRCRBu}.


\section{System Model}
\label{sec:system_model}

We consider a multi-user uplink scenario where a \ac{BS} equipped with an $N$-element \ac{ULA} receives $L$ snapshots from $K$ \ac{NF} users. The channel of each user is modeled as a superposition of multiple propagation paths, and the spacing of the elements is $\Delta=\lambda/2$. The position of the antenna element $n$ is $x_n=\delta_n\Delta$, where $\delta_n=n-(N+1)/2$, $n=1,\ldots,N$.

\begin{figure}[t]
    \centering

\begin{tikzpicture}[font=\footnotesize, line cap=round, line join=round]

\def\R{4.1}
\def\rInner{1.0}
\def\yShift{0.35}

\definecolor{ringgray}{RGB}{110,110,110}
\definecolor{nfregion}{RGB}{205,230,205}
\definecolor{bscolor}{RGB}{0,85,145}
\definecolor{nloscolor}{RGB}{190,60,60}
\definecolor{loscolor}{RGB}{0,135,85}
\definecolor{wificolor}{RGB}{95,95,95}

\newcommand{\UEphone}[5]{%
  \begin{scope}
    \draw[fill=black, line width=0.5pt, rounded corners=0.6pt]
      (#1,#2) rectangle ++(0.24,0.42);
    \fill[gray!25] (#1+0.03,#2+0.10) rectangle ++(0.18,0.20);
    \fill[gray!55] (#1+0.12,#2+0.06) circle (0.012);
    \node[above=2pt] at (#1+0.12+#4,#2+0.44+#5) {#3};
  \end{scope}
}

\newcommand{\UplinkWavesToBS}[2]{%
  \begin{scope}
    \coordinate (UEc) at (#1,#2);
    \pgfmathanglebetweenpoints{\pgfpointanchor{UEc}{center}}{\pgfpointanchor{BSref}{center}}
    \let\angToBS\pgfmathresult
    \begin{scope}[shift={(UEc)}, rotate=\angToBS]
      \foreach \r in {0.18,0.28,0.38}{%
        \draw[wificolor, line width=0.55pt]
          ({\r*cos(-30)},{\r*sin(-30)})
          arc[start angle=-30, end angle=30, radius=\r];
      }
    \end{scope}
  \end{scope}
}

\newcommand{\ULAdraw}[4]{%
  \begin{scope}
    \draw[fill=bscolor, line width=0.8pt] (#1,#2) -- ++(#3,0);
    \pgfmathsetmacro{\dx}{#3/(#4-1)}

    \pgfmathsetmacro{\dotR}{max(0.010,0.16*\dx)}
    \pgfmathsetmacro{\stem}{max(0.10,7.40*\dotR)}
    \pgfmathsetmacro{\armW}{0.32*\dx}
    \pgfmathsetmacro{\armH}{max(0.08,1.10*\dotR)}
    \pgfmathsetmacro{\dotsY}{max(0.12,2.2*\dotR)}

    \pgfmathsetmacro{\nEdgeReal}{min(4,max(2,floor(#3/(0.55))))}
    \pgfmathtruncatemacro{\nEdge}{\nEdgeReal}
    \pgfmathtruncatemacro{\iLeftLast}{\nEdge-1}
    \pgfmathtruncatemacro{\iRightStart}{#4-\nEdge}

    \newcommand{\drawElem}[1]{%
      \fill[bscolor] (#1+##1*\dx,#2) circle (\dotR);
      \draw[bscolor,line width=0.7pt] (#1+##1*\dx,#2) -- ++(0,\stem);
      \draw[bscolor,line width=0.8pt]
        (#1+##1*\dx,#2+\stem) -- ++(-\armW,\armH);
      \draw[bscolor,line width=0.8pt]
        (#1+##1*\dx,#2+\stem) -- ++(\armW,\armH);
    }

    \foreach \i in {0,...,\iLeftLast} { \drawElem{\i} }
    \foreach \i in {\iRightStart,...,\numexpr#4-1\relax} { \drawElem{\i} }

    \node at (#1+0.5*#3,#2+\dotsY) {$\cdots$};
    \node[below=2pt] at (#1+0.5*#3,#2) {};
  \end{scope}
}
\draw[ringgray,line width=0.7pt] (-\R,\yShift) arc (180:360:\R);
\fill[nfregion] (-\rInner,\yShift) arc (180:360:\rInner) -- (\rInner,\yShift) -- cycle;

\coordinate (InnerBoundR) at (\rInner,\yShift);
\coordinate (OuterBoundR) at (\R,\yShift);

\fill[black] (InnerBoundR) circle (0.8pt);
\fill[black] (OuterBoundR) circle (0.8pt);

\node[right=3pt,xshift=-20pt,yshift=6pt] at (OuterBoundR) {$(R_F,0)$};
\node[right=3pt,xshift=-4pt,yshift=6pt] at (InnerBoundR) {$(2D,0)$};

\fill[bscolor] (0,\yShift) circle (0.06);
\node[align=center] at (0,\yShift+0.6) {BS\\$(0,0)$};
\coordinate (BSref) at (0,\yShift+0.05);

\UEphone{-3.25}{\yShift-1.0}{user 1}{0}{0}
\UplinkWavesToBS{-3.20+0.12}{\yShift-0.9+0.21}

\UEphone{-2.5}{\yShift-2.55}{user 2}{0}{0}
\UplinkWavesToBS{-2.45+0.12}{\yShift-2.5+0.21}

\UEphone{-0.30}{\yShift-3.20}{}{0}{0}
\node[right=4pt] at (-0.2,\yShift-2.89) {user 3};

\UplinkWavesToBS{-0.20}{\yShift-3.10+0.21}
\coordinate (UE3) at (-0.20,\yShift-3.10+0.21);

\node at (1.25,\yShift-2.56) {$\cdots$};

\def\UEKx{2.65}
\def\UEKy{\yShift-2.5}
\UEphone{\UEKx}{\UEKy}{user $K$}{0.40}{0.00}
\UplinkWavesToBS{\UEKx+0.12}{\UEKy+0.21}

\coordinate (UEK) at (\UEKx+0.12,\UEKy+0.21);

\coordinate (ScL1) at (2.6,\yShift-1.55);
\coordinate (ScR2) at (2.1,\yShift-2.45);

\fill[nloscolor] (ScL1) circle (0.055);
\fill[nloscolor] (ScR2) circle (0.055);

\coordinate (Sc3L1) at (-0.75,\yShift-2.35);
\coordinate (Sc3R2) at (0.38,\yShift-2.35);

\fill[nloscolor] (Sc3L1) circle (0.055);
\fill[nloscolor] (Sc3R2) circle (0.055);

\tikzset{
  nlosray/.style={nloscolor,dashed,line width=1pt},
  nlosarrow/.style={nloscolor,dashed,line width=1pt,-{Stealth[length=3pt]}},
  lospath/.style={loscolor,line width=1.25pt,-{Stealth[length=3pt]}}
}

\draw[nlosray]   (UEK) -- (ScL1);
\draw[nlosarrow] (ScL1) -- (BSref);

\draw[nlosray]   (UEK) -- (ScR2);
\draw[nlosarrow] (ScR2) -- (BSref);

\draw[nlosray]   (UE3) -- (Sc3L1);
\draw[nlosarrow] (Sc3L1) -- (BSref);

\draw[nlosray]   (UE3) -- (Sc3R2);
\draw[nlosarrow] (Sc3R2) -- (BSref);

\draw[lospath] (UEK) -- (BSref);
\draw[lospath] (UE3) -- (BSref);
\node[
loscolor,
fill=white,
inner sep=0.01pt,
font=\tiny

] 
at (1.55,\yShift-1.2) {$d_k^{(1)}$};

\draw[loscolor, line width=0.5pt]
(1.22,\yShift) arc[start angle=0,end angle=-22,radius=2.32];

\node[
loscolor,
fill=white,
inner sep=0.1pt,
font=\tiny

] 
at (0.9,\yShift-0.33) {$\phi_k^{(1)}$};

\ULAdraw{-0.9}{\yShift+0.02}{1.8}{9}

\end{tikzpicture}
\caption{\ac{NF} multi-user uplink system. Users are located in the radiative \ac{NF} region, i.e., between $2D$ and $R_F$, where $D=(N-1)\lambda/2$ and $R_F=2D^2/\lambda$. For the $p$-th path of user $k$, $d_k^{(p)}$ denotes the path range from the array reference point, and $\phi_k^{(p)}$ is the corresponding azimuth angle measured with respect to the array axis.}
\end{figure}
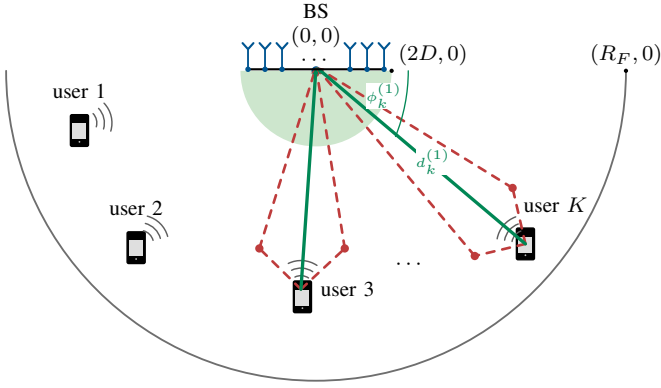

The $p$-th path of user $k$ is characterized by $\bm{\theta}_k^{(p)}=(d_k^{(p)},\phi_k^{(p)})$, for $p=1,\ldots,P_k$, where $d_k^{(p)}$, $\phi_k^{(p)}$, and $P_k$ denote the range, azimuth, and number of propagation paths of user $k$, respectively. Under the spherical-wave model, the distance between antenna $n$ and path $p$ of user $k$ is
\begin{equation}
r_{n,k}^{(p)}
=
\sqrt{\left(d_k^{(p)}\right)^2+(\delta_n\Delta)^2
-2d_k^{(p)}\delta_n\Delta\cos\!\left(\phi_k^{(p)}\right)}.
\label{eq:distance_path}
\end{equation}

The corresponding \ac{NF} steering vector associated with the parameter pair $\bm{\theta}_k^{(p)}$ is defined as
\begin{equation}
\mathbf{h}\!\left(\bm{\theta}_k^{(p)}\right)
=
\left[
e^{-j\frac{2\pi}{\lambda}r_{1,k}^{(p)}},
\ldots,
e^{-j\frac{2\pi}{\lambda}r_{N,k}^{(p)}}
\right]^{\mathsf T}
\in\mathbb{C}^{N\times1}.
\label{eq:steering_path}
\end{equation}

The effective channel vector of user $k$ is modeled as the superposition of $P_k$ propagation paths \cite{8269397}, i.e.,
\begin{equation}
\mathbf{h}_k
=
\sum_{p=1}^{P_k}\alpha_k^{(p)}
\mathbf{h}\!\left(\bm{\theta}_k^{(p)}\right),
\qquad
\mathbf{h}_k\in\mathbb{C}^{N\times1},
\label{eq:user_channel}
\end{equation}
where $\alpha_k^{(p)}\in\mathbb{C}$ denotes the complex gain of the $p$-th propagation path. Any user-specific phase offset can be absorbed into the complex path gains $\alpha_k^{(p)}$ and is therefore omitted for notational simplicity. Stacking the channels of all users yields the multi-user channel matrix $\mathbf{H}=[\mathbf{h}_1,\ldots,\mathbf{h}_K]\in\mathbb{C}^{N\times K}$.
Let $\mathbf{s}[\ell]\in\mathbb{C}^{K\times1}$ denote the transmitted symbol vector at snapshot $\ell$, and let $\mathbf{y}[\ell]\in\mathbb{C}^{N\times1}$ denote the received signal vector at the BS. The received signal model is given by
\begin{equation}
\mathbf{y}[\ell]
=
\sqrt{P_s}\,\mathbf{H}\mathbf{s}[\ell]+\mathbf{w}[\ell],
\qquad \ell=1,\ldots,L,
\label{eq:received_snapshot}
\end{equation}
where $P_s$ denotes the transmit power, and $\mathbf{w}[\ell] \sim \mathcal{CN}(\mathbf{0},\sigma^2\mathbf{I}_N)$ represents additive white Gaussian noise with variance $\sigma^2$. Collecting all snapshots gives
$\mathbf{Y}=[\mathbf{y}[1],\ldots,\mathbf{y}[L]]\in\mathbb{C}^{N\times L}$. The transmitted symbols are i.i.d. circularly symmetric complex Gaussian.


\section{PR-Based Multipath Channel Estimation}
\label{sec:pr}

This section presents the proposed \ac{PR}-based formulations for \ac{NF} multipath channel estimation in the known-pilot and unknown-symbol cases, building on the \ac{PR} framework in \cite{AbuAqoulahICC}.

\paragraph{General \ac{PR} formulation}

\ac{PR} estimates the desired component parametrically while absorbing the remaining interfering components into an unstructured nuisance term \cite{TrinhHoang2018PRFramework}.
In the proposed \ac{NF} multipath uplink setting, we apply this principle to estimate the desired-user multipath channel while relaxing the remaining multi-user interference.

The desired user's channel is represented as a superposition of $P_1$ propagation paths, where $P_1$ is not assumed known a priori. Let $\bm{\Theta}_1=\{\bm{\theta}_1^{(p)}\}_{p=1}^{P_1}$ denote its path-parameter set, with $\bm{\theta}_1^{(p)}=(d_1^{(p)},\phi_1^{(p)})$, and let $\mathbf{a}_1=[\alpha_1^{(1)},\ldots,\alpha_1^{(P_1)}]^{\mathsf T}\in\mathbb{C}^{P_1\times1}$ collect the corresponding path gains. The resulting multipath steering matrix is
\begin{equation}
\mathbf{H}(\bm{\Theta}_1)=\big[\mathbf{h}(\bm{\theta}_1^{(1)})\ \cdots\ \mathbf{h}(\bm{\theta}_1^{(P_1)})\big]\in\mathbb{C}^{N\times P_1}.
\end{equation}
Let $\mathbf{s}_1\in\mathbb{C}^{L\times1}$ denote the signal sequence of the desired user, and $\mathbf{S}_B=[\mathbf{s}_2\ \cdots\ \mathbf{s}_K]\in\mathbb{C}^{L\times(K-1)}$ collect the signal sequences of the other users. The resulting \ac{PR} data model is
\begin{equation}
\mathbf{Y}
=
\underbrace{\mathbf{H}(\bm{\Theta}_1)\mathbf{a}_1\mathbf{s}_1^{\mathsf T}}_{\text{Desired user}}
+
\underbrace{\mathbf{B}\mathbf{S}_B^{\mathsf T}}_{\text{Other users (relaxed)}}
+
\underbrace{\mathbf{W}}_{\text{Noise}}.
\end{equation}
where $\mathbf{B}\in\mathbb{C}^{N\times(K-1)}$ is an unstructured nuisance matrix representing the aggregate contribution of the remaining users under the \ac{PR} formulation, with the transmit-power scaling and per-user phase offsets absorbed into $\mathbf{a}_1$ for the desired user and into $\mathbf{B}$ for the remaining users.
\begin{equation}
J(\bm{\Theta}_1,\mathbf{a}_1,\mathbf{B})
=
\left\|
\mathbf{Y}-\mathbf{H}(\bm{\Theta}_1)\mathbf{a}_1\mathbf{s}_1^{\mathsf T}-\mathbf{B}\mathbf{S}_B^{\mathsf T}
\right\|_F^2 .
\end{equation}
Depending on whether $\mathbf{s}_1$ and $\mathbf{S}_B$ are known or unknown, this model yields a multipath \ac{PR-ML} formulation for the known-pilot case and a rank-adaptive multipath \ac{PR-CF} formulation for the unknown-symbol case. The proposed \ac{PR}-based estimators reduce complexity by sequentially estimating the desired user's dominant paths and absorbing the remaining users into a relaxed nuisance term, while achieving better performance than the \ac{2D-MUSIC} baseline at comparable complexity.

\paragraph{\ac{PR-ML} with known pilots}

For known pilots, we exploit the knowledge of $\mathbf{s}_1$ and $\mathbf{S}_B$ and develop a greedy procedure to estimate the desired user's path parameters. Minimizing the cost with respect to the matrix $\mathbf{B}$ yields
\begin{equation}
J(\bm{\Theta}_1,\mathbf{a}_1)
=
\left\|
\mathbf{R}(\bm{\Theta}_1,\mathbf{a}_1)\mathbf{P}_{\mathbf{S}_B}^{\perp}
\right\|_F^2 ,
\end{equation}
where $\mathbf{R}(\bm{\Theta}_1,\mathbf{a}_1)=\mathbf{Y}-\mathbf{H}(\bm{\Theta}_1)\mathbf{a}_1\mathbf{s}_1^{\mathsf T}$ and, assuming that $\mathbf{S}_B$ has full column rank, $\mathbf{P}_{\mathbf{S}_B}^{\perp}=\mathbf{I}_L-\mathbf{S}_B(\mathbf{S}_B^{\mathsf H}\mathbf{S}_B)^{-1}\mathbf{S}_B^{\mathsf H}$. The projected desired-user pilot must satisfy $\mathbf{P}_{\mathbf{S}_B}^{\perp}\mathbf{s}_1\neq\mathbf{0}$, which holds for orthogonal or sufficiently well-conditioned pilot sequences. Further concentrating with respect to $\mathbf{a}_1$, and defining $\tilde{\mathbf{Y}}=\mathbf{Y}\mathbf{P}_{\mathbf{S}_B}^{\perp}$ and $\tilde{\mathbf{s}}_1=\mathbf{P}_{\mathbf{S}_B}^{\perp}\mathbf{s}_1$, we obtain the compressed observation $\bar{\mathbf{y}}=\tilde{\mathbf{Y}}\tilde{\mathbf{s}}_1^{*}/\|\tilde{\mathbf{s}}_1\|_2^2$. This is the matched-filtered spatial observation after nuisance-pilot projection, and it preserves the least-squares solution for the desired-user channel. The LS estimate of $\mathbf{a}_1$ then leads to the multipath \ac{PR-ML} score
\begin{equation}
S_{\mathrm{PR\text{-}ML}}(\bm{\Theta}_1)
=
\bar{\mathbf{y}}^{\mathsf H}
\bm{\Pi}_{\mathbf{H}(\bm{\Theta}_1)}
\bar{\mathbf{y}},
\end{equation}
where $\bm{\Pi}_{\mathbf{H}(\bm{\Theta}_1)}=\mathbf{H}(\bm{\Theta}_1)\big(\mathbf{H}^{\mathsf H}(\bm{\Theta}_1)\mathbf{H}(\bm{\Theta}_1)\big)^{-1}\mathbf{H}^{\mathsf H}(\bm{\Theta}_1)$. We then employ a greedy path extraction procedure that sequentially identifies dominant paths and updates their gains, as summarized in Algorithm~\ref{alg:prml_mp_auto}. The estimated number of paths is given by the final greedy iteration index, and therefore it may differ from the true number of physical paths when weak paths are below the noise floor or when noise/interference components are over-extracted. The stopping threshold $\varepsilon$ and the upper bound $P_{\max}$ are used to control this underestimation--overestimation tradeoff.

\begin{algorithm}[t]
\caption{\ac{PR-ML} Multipath Estimation}
\label{alg:prml_mp_auto}

\textbf{Input:}
Received data $\mathbf{Y}$, pilot $\mathbf{s}_1$, pilot matrix $\mathbf{S}_B$, 
search grid $\mathcal{G}$, maximum number of paths $P_{\max}$, threshold $\epsilon$.

\textbf{Output:}
Estimated path set $\hat{\bm{\Theta}}_1$, gains $\hat{\mathbf{a}}_1$, channel $\hat{\mathbf{h}}_1$.

1: Compute projection matrix  
$\mathbf{P}_{\mathbf{S}_B}^{\perp} = \mathbf{I}_L-\mathbf{S}_B(\mathbf{S}_B^{\mathsf H}\mathbf{S}_B)^{-1}\mathbf{S}_B^{\mathsf H}$

2: Project received data  
$\tilde{\mathbf{Y}}=\mathbf{Y}\mathbf{P}_{\mathbf{S}_B}^{\perp}$,\quad
$\tilde{\mathbf{s}}_1=\mathbf{P}_{\mathbf{S}_B}^{\perp}\mathbf{s}_1$

3: Compress snapshots  
$\bar{\mathbf{y}}=\tilde{\mathbf{Y}}\tilde{\mathbf{s}}_1^{*}/\|\tilde{\mathbf{s}}_1\|_2^2$

4: Initialize  
$p=0$, $\mathbf{r}^{(0)}=\bar{\mathbf{y}}$, $\mathbf{H}^{(0)}=[\ ]$

5: \textbf{repeat}

6:\hspace{0.4cm} $p=p+1$

7:\hspace{0.4cm} Select path  
$\hat{\bm{\theta}}_1^{(p)}=
\arg\max_{\bm{\theta}\in\mathcal G}
\frac{|\mathbf{h}(\bm{\theta})^{\mathsf H}\mathbf{r}^{(p-1)}|^2}{\|\mathbf{h}(\bm{\theta})\|_2^2}$

8:\hspace{0.4cm} Update steering matrix  
$\mathbf{H}^{(p)}=[\mathbf{H}^{(p-1)}\ \mathbf{h}(\hat{\bm{\theta}}_1^{(p)})]$

9:\hspace{0.4cm} Estimate gains  
$\hat{\mathbf{a}}_1^{(p)}=(\mathbf{H}^{(p)\mathsf H}\mathbf{H}^{(p)})^{-1}\mathbf{H}^{(p)\mathsf H}\bar{\mathbf{y}}$

10:\hspace{0.4cm} Update residual  
$\mathbf{r}^{(p)}=\bar{\mathbf{y}}-\mathbf{H}^{(p)}\hat{\mathbf{a}}_1^{(p)}$

11: \textbf{until} $\|\mathbf{r}^{(p)}\|_2^2/\|\mathbf{r}^{(0)}\|_2^2\le\epsilon$ 
or $p=P_{\max}$

12: Set $\hat{P}_1=p$,
$\hat{\bm{\Theta}}_1=\{\hat{\bm{\theta}}_1^{(p)}\}_{p=1}^{\hat{P}_1}$,
$\hat{\mathbf{a}}_1=\hat{\mathbf{a}}_1^{(\hat{P}_1)}$,
$\hat{\mathbf{h}}_1=\mathbf{H}^{(\hat{P}_1)}\hat{\mathbf{a}}_1$.

\end{algorithm}


\paragraph{\ac{PR-CF} with unknown symbols}

For unknown symbols, we adopt a rank-adaptive multipath \ac{PR-CF} formulation based on the sample covariance matrix $\mathbf{R}_b\triangleq\mathbf{Y}\mathbf{Y}^{\mathsf H}\in\mathbb{C}^{N\times N}$. Building on the single-path \ac{PR-CF} model in \cite{jade_pr_paper} and the near-field single-path formulation in \cite{AbuAqoulahICC}, we consider a discretized range--azimuth grid $\mathcal{G}\subset\mathbb{R}_+\times\mathbb{R}$. Here, $\mathbf{h}(\bm{\theta})$ denotes the \ac{NF} steering vector associated with a generic parameter pair $\bm{\theta}=(d,\phi)$, as defined in Section~\ref{sec:system_model}. We further assume that the dominant paths of different users lie in disjoint user-specific regions $\{\mathcal{G}_k\}_{k=1}^K$, with $\mathcal{G}_k\cap\mathcal{G}_\ell=\emptyset$ for $k\neq\ell$, so that the greedy search targets the multipath structure of a single user. Following the \ac{PR-CF} formulation, the score of a candidate multipath steering matrix is computed from the matrices below. As in covariance-fitting PR methods, we assume that $\mathbf{R}_b$ is positive definite. This typically holds when $L\geq N$ and the observations are sufficiently rich. If $\mathbf{R}_b$ is singular or ill-conditioned, diagonal loading can be applied \cite{TrinhHoang2018PRFramework,Li2003DiagonalLoading}. Define
\begin{equation}
\mathbf{T}\!\left(\mathbf{H}^{(p)}\right)
=
\mathbf{H}^{(p)}
\left(
\mathbf{H}^{(p)\mathsf H}\mathbf{R}_b^{-1}\mathbf{H}^{(p)}
\right)^{-1}
\mathbf{H}^{(p)\mathsf H},
\end{equation}
and
\begin{equation}
\mathbf{M}\!\left(\mathbf{H}^{(p)}\right)
=
\mathbf{R}_b-\mathbf{T}\!\left(\mathbf{H}^{(p)}\right).
\end{equation}
Let $\lambda_i(\cdot)$ denote the $i$-th largest eigenvalue. The resulting rank-adaptive multipath \ac{PR-CF} objective is
\begin{equation}
f_{\mathrm{PR\text{-}CF}}\!\left(\mathbf{H}^{(p)}\right)
=
\frac{1}{
\sum_{i=K}^{N}
\lambda_i^2\!\left(
\mathbf{M}\!\left(\mathbf{H}^{(p)}\right)
\right)
}.
\end{equation}
The eigenvalue-based objective follows from the low-rank covariance-fitting interpretation of \ac{PR}: for a given candidate desired-user subspace, the relaxed nuisance component is represented by the dominant low-rank part of the residual covariance, while the remaining fitting error is captured by the sum of the squared residual eigenvalues \cite{TrinhHoang2018PRFramework}.

The dominant paths are identified through a greedy search over the user-specific region $\mathcal{G}_k$. Starting from $\mathbf{H}^{(0)}=\emptyset$, each candidate location $\bm{\theta}\in\mathcal{G}_k$ defines the temporary steering matrix
\begin{equation}
\mathbf{H}^{(p)}(\bm{\theta})
=
\big[
\mathbf{H}^{(p-1)}\ \mathbf{h}(\bm{\theta})
\big],
\end{equation}
and the next path is selected as
\begin{equation}
\hat{\bm{\theta}}^{(p)}
=
\arg\max_{\bm{\theta}\in\mathcal{G}_k}
f_{\mathrm{PR\text{-}CF}}\!\left(\mathbf{H}^{(p)}(\bm{\theta})\right).
\end{equation}
After selecting $p$ paths, the steering matrix is updated as
\begin{equation}
\mathbf{H}^{(p)}
=
\big[
\mathbf{h}(\hat{\bm{\theta}}^{(1)})
\ \cdots\
\mathbf{h}(\hat{\bm{\theta}}^{(p)})
\big]
\in\mathbb{C}^{N\times p}.
\end{equation}
The first path is always extracted. From the second iteration onward, the procedure stops when the relative improvement in the objective falls below $\varepsilon$ or when $p=P_{\max}$. The complete estimation procedure is summarized in Algorithm~\ref{alg:prcf_nlos}.

\begin{algorithm}[t]
\caption{Greedy PR-CF Multipath Estimation}
\label{alg:prcf_nlos}

\textbf{Input:}  
Received data $\mathbf{Y}$, user-specific search region $\mathcal{G}_k \subseteq \mathcal{G}$, maximum number of paths $P_{\max}$, threshold $\varepsilon$.

\textbf{Output:}  
Estimated path set $\hat{\bm{\Theta}}$, channel $\hat{\mathbf h}$.

1: Compute sample covariance  
$\mathbf{R}_b=\mathbf{Y}\mathbf{Y}^{\mathsf H}$

2: Initialize  
$p=0$, $\mathbf{H}^{(0)}=\emptyset$, $\bm{\Theta}^{(0)}=\emptyset$, $f^{(0)}=0$

3: \textbf{repeat}

4:\hspace{0.4cm} $p=p+1$

5:\hspace{0.4cm} For each candidate $\bm{\theta}\in\mathcal{G}_k$

6:\hspace{0.8cm} Form steering matrix  
$\mathbf{H}^{(p)}(\bm{\theta})=[\mathbf{H}^{(p-1)}\ \mathbf{h}(\bm{\theta})]$

7:\hspace{0.8cm} Compute  
$\mathbf{T}\!\left(\mathbf{H}^{(p)}(\bm{\theta})\right)$

8:\hspace{0.8cm} Compute  
$\mathbf{M}\!\left(\mathbf{H}^{(p)}(\bm{\theta})\right)
=
\mathbf{R}_b-\mathbf{T}\!\left(\mathbf{H}^{(p)}(\bm{\theta})\right)$

9:\hspace{0.8cm} Evaluate score  
$f_{\mathrm{PR\text{-}CF}}\!\left(\mathbf{H}^{(p)}(\bm{\theta})\right)$

10:\hspace{0.4cm} Select path  
$\hat{\bm{\theta}}^{(p)}
=
\arg\max_{\bm{\theta}\in\mathcal{G}_k}
f_{\mathrm{PR\text{-}CF}}\!\left(\mathbf{H}^{(p)}(\bm{\theta})\right)$

11:\hspace{0.4cm} Update steering matrix  
$\mathbf{H}^{(p)}=[\mathbf{H}^{(p-1)}\ \mathbf{h}(\hat{\bm{\theta}}^{(p)})]$

12:\hspace{0.4cm} Update path set  
$\bm{\Theta}^{(p)}=\bm{\Theta}^{(p-1)}\cup\{\hat{\bm{\theta}}^{(p)}\}$

13:\hspace{0.4cm} Update score  
$f^{(p)}=f_{\mathrm{PR\text{-}CF}}(\mathbf{H}^{(p)})$

14: Compute $\Delta f^{(p)}=\frac{f^{(p)}-f^{(p-1)}}{f^{(p-1)}}$.

15: until $(p \geq 2 \ \text{AND}\ \Delta f^{(p)} < \epsilon)$ or $p=P_{\max}$

16: Set $p_{\mathrm{out}}=p-1$ if ($p \geq 2$ AND $\Delta f^{(p)}<\epsilon$); otherwise,

set $p_{\mathrm{out}}=p$.

17: Set $\hat{\bm{\Theta}}=\bm{\Theta}^{(p_{\mathrm{out}})}$.

18: Form the corresponding steering matrix and reconstruct $\hat{\mathbf{h}}$.

\end{algorithm}


\paragraph{\ac{2D-MUSIC} Baseline}
As a baseline, we consider \ac{NF} \ac{2D-MUSIC} \cite{gurgunoglu2025_2dmusic}, where each propagation path is treated as an independent source and the channel is reconstructed from the detected peaks. The pseudospectrum is
\begin{equation}
S_{\mathrm{MUSIC}}(\bm{\theta})=
\frac{1}{\mathbf{h}(\bm{\theta})^{\mathsf H}\hat{\mathbf{U}}_n\hat{\mathbf{U}}_n^{\mathsf H}\mathbf{h}(\bm{\theta})},
\end{equation}
where $\hat{\mathbf{U}}_n$ is the noise subspace of $\mathbf{R}_b$. Evaluating $S_{\mathrm{MUSIC}}(\bm{\theta})$ over $\mathcal{G}$, dominant peaks are extracted sequentially using $S_{\mathrm{MUSIC}}(\hat{\bm{\theta}})\geq\varepsilon S_{\max}$, where $S_{\max}=\max_{\bm{\theta}\in\mathcal{G}} S_{\mathrm{MUSIC}}(\bm{\theta})$, followed by \ac{NMS}. Let $\hat Q=|\hat{\bm{\Theta}}|$ with $\hat{\bm{\Theta}}=\{\hat{\bm{\theta}}_q\}_{q=1}^{\hat Q}$, and form $\hat{\mathbf{H}}=[\mathbf{h}(\hat{\bm{\theta}}_1)\ \cdots\ \mathbf{h}(\hat{\bm{\theta}}_{\hat Q})]$. The path gains are then estimated as $\hat{\mathbf{a}}=(\hat{\mathbf{H}}^{\mathsf H}\hat{\mathbf{H}})^{-1}\hat{\mathbf{H}}^{\mathsf H}\bar{\mathbf{y}}$.

\section{Cram\'er--Rao Bounds for Partial Relaxation}
\label{sec:crb}
This section derives multipath \acp{CRB} under the \ac{PR} framework for \ac{NF} channel estimation. In particular, we develop \ac{PRCRBp} for the known-pilot case and \ac{PRCRBu} for the unknown-symbol case.

\paragraph{\ac{PRCRBp} for known pilots}

For the known-pilot case, we derive the multipath \ac{PRCRBp} under the \ac{PR} framework. The received data follow
\begin{equation}
\mathbf{Y}
=
\mathbf{H}(\bm{\Theta}_1)\mathbf{a}_1\mathbf{s}_1^{\mathsf T}
+
\mathbf{B}\mathbf{S}_B^{\mathsf T}
+
\mathbf{W},
\end{equation}
Let $\mathbf{y}=\mathrm{vec}(\mathbf{Y})\in\mathbb{C}^{NL}$ and $\mathbf{b}=\mathrm{vec}(\mathbf{B})\in\mathbb{C}^{N(K-1)}$. Then
\begin{equation}
\mathbf{y}
=
\underbrace{
(\mathbf{s}_1\otimes\mathbf{I}_N)\mathbf{H}(\bm{\Theta}_1)\mathbf{a}_1
+
(\mathbf{S}_B\otimes\mathbf{I}_N)\mathbf{b}
}_{\triangleq\mathbf{m}(\bm{\eta})}
+
\mathbf{w},
\end{equation}
where $\mathbf{m}\triangleq\mathbf{m}(\bm{\eta})\in\mathbb{C}^{NL}$ is the mean of $\mathbf{y}$, and $\bm{\eta}=[\bm{\vartheta}^{\mathsf T},\bm{\xi}^{\mathsf T}]^{\mathsf T}\in\mathbb{R}^{4P_1+2N(K-1)}$, with $\bm{\vartheta}\in\mathbb{R}^{4P_1}$ and $\bm{\xi}\in\mathbb{R}^{2N(K-1)}$. Following the real-valued CRB parametrization for complex-valued array models in \cite{Stoica1990ConditionalUnconditional}, we collect the real physical parameters together with the real and imaginary parts of the complex coefficients. Hence,
\[
\bm{\vartheta}
=
[\mathbf{d}^{\mathsf T},\bm{\phi}^{\mathsf T},\Re\{\mathbf{a}_1\}^{\mathsf T},\Im\{\mathbf{a}_1\}^{\mathsf T}]^{\mathsf T},
\quad
\bm{\xi}
=
[\Re\{\mathbf{b}\}^{\mathsf T},\Im\{\mathbf{b}\}^{\mathsf T}]^{\mathsf T}.
\]
where $\mathbf{d},\bm{\phi}\in\mathbb{R}^{P_1}$. Using the general Gaussian FIM in \cite[Sec.~3.9, Eq.~(3.31)]{kay1993}, 
and noting that the covariance $\sigma^2\mathbf{I}$ is parameter independent, 
the FIM for the real-valued parameter vector $\bm{\eta}$ becomes
\begin{equation*}
\mathbf{F}(\bm{\eta})=
\frac{2}{\sigma^2}
\Re\!\left\{
\left(
\frac{\partial\mathbf{m}}{\partial\bm{\eta}}
\right)^{\!\mathsf H}
\left(
\frac{\partial\mathbf{m}}{\partial\bm{\eta}}
\right)
\right\},
\quad
\mathbf{F}(\bm{\eta})=
\begin{bmatrix}
\mathbf{J}_{\vartheta\vartheta} & \mathbf{J}_{\vartheta\xi}\\
\mathbf{J}_{\xi\vartheta} & \mathbf{J}_{\xi\xi}
\end{bmatrix}.
\end{equation*}
Eliminating the nuisance parameter vector $\bm{\xi}$ via the Schur complement yields $\mathbf{F}_{\mathrm{eff}}=\mathbf{J}_{\vartheta\vartheta}-\mathbf{J}_{\vartheta\xi}\mathbf{J}_{\xi\xi}^{-1}\mathbf{J}_{\xi\vartheta}$, so that $\mathrm{PR\text{-}CRB}_p(\bm{\vartheta})=\mathbf{F}_{\mathrm{eff}}^{-1}$. Let $\mathbf{h}_p=\mathbf{h}(\bm{\theta}_{1}^{(p)})$ and $\mathbf{S}=\mathbf{s}_1\otimes\mathbf{I}_N$. The required Jacobian terms are
\begin{equation*}
\begin{alignedat}{2}
\frac{\partial\mathbf{m}}{\partial\Re\{\alpha_{1}^{(p)}\}} &= \mathbf{S}\mathbf{h}_p, \qquad&
\frac{\partial\mathbf{m}}{\partial d_{1}^{(p)}} &= \mathbf{S}\,\alpha_{1}^{(p)}\frac{\partial\mathbf{h}_p}{\partial d_{1}^{(p)}},\\
\frac{\partial\mathbf{m}}{\partial\Im\{\alpha_{1}^{(p)}\}} &= j\,\mathbf{S}\mathbf{h}_p, \qquad&
\frac{\partial\mathbf{m}}{\partial \phi_{1}^{(p)}} &= \mathbf{S}\,\alpha_{1}^{(p)}\frac{\partial\mathbf{h}_p}{\partial \phi_{1}^{(p)}}.
\end{alignedat}
\end{equation*}
The derivatives of $\mathbf{h}_p$ follow from \cite{gurgunoglu2025_2dmusic}. Let $\mathbf{J}_h=\partial(\mathbf{H}(\bm{\Theta}_1)\mathbf{a}_1)/\partial\bm{\vartheta}^{\mathsf T}$. The channel-domain bound is
\begin{equation}
\mathbf{CRB}_{\mathbf{h}}
=
\mathbf{J}_h\,
\mathbf{PR\text{-}CRB}_p(\bm{\vartheta})\,
\mathbf{J}_h^{\mathsf T}.
\end{equation}


\paragraph{\ac{PRCRBu} for unknown symbols}
Based on the constrained \ac{PRCRBu} formulation in \cite{AbuAqoulahICC,stoica1998ccrb}, we derive the multipath \ac{PRCRBu} for the unknown-symbol case within the \ac{PR} framework, as a benchmark for the multipath \ac{PR-CF} estimator, where the desired-user channel is modeled by $\mathbf{H}(\bm{\Theta}_1)\mathbf{a}_1$. In this setting, we represent the relaxed interference block $\mathbf{B}\mathbf{S}_B^{\mathsf T}$ through its singular-value decomposition (SVD) as $\mathbf{U}_r\bm{\Gamma}\mathbf{V}_r^{\mathsf H}$. The resulting received signal model is
\begin{equation}
\mathbf{Y}
=
\mathbf{H}(\bm{\Theta}_1)\mathbf{a}_1\mathbf{s}_1^{\mathsf T}
+
\mathbf{U}_r\bm{\Gamma}\mathbf{V}_r^{\mathsf H}
+
\mathbf{W}.
\end{equation}
Here, $r$ denotes the rank of the relaxed interference model, $\mathbf{U}_r=[\mathbf{u}_1,\ldots,\mathbf{u}_r]\in\mathbb{C}^{N\times r}$, $\mathbf{V}_r=[\mathbf{v}_1,\ldots,\mathbf{v}_r]\in\mathbb{C}^{L\times r}$, and $\bm{\Gamma}=\mathrm{diag}(\bm{\gamma})\in\mathbb{R}_+^{r\times r}$ is a diagonal matrix containing the singular values. In addition to the non-uniqueness of the relaxed-block factorization, the desired-user term $\mathbf{H}(\bm{\Theta}_1)\mathbf{a}_1\mathbf{s}_1^{\mathsf T}$ is identifiable only up to a common complex scale/phase factor between $\mathbf{a}_1$ and $\mathbf{s}_1$. Without loss of generality, a set of constraints is defined to make the factorizations unique, and hence to obtain a finite bound. Specifically, the columns of $\mathbf{U}_r$ and $\mathbf{V}_r$ are constrained to having a unit norm, and the first element of each vector $\mathbf{u}_k$ is set to be real because any phase rotation can be included in the corresponding vector $\mathbf{v}_k$. As a result, the set of constraints is $\|\mathbf{u}_k\|_2=\|\mathbf{v}_k\|_2=1$ and $\Im\{[\mathbf{u}_k]_1\}=0$, $\forall k$, and $\Re\{[\mathbf{s}_1]_1\}=1$, $\Im\{[\mathbf{s}_1]_1\}=0$. 

Using the same vectorized model and Fisher-information matrix as in the known-pilot case, the mean vector of $\mathbf{y}$ is
\begin{equation}
\bm{\mu}(\bm{\zeta})
=
(\mathbf{s}_1\otimes\mathbf{I}_N)\mathbf{H}(\bm{\Theta}_1)\mathbf{a}_1
+
\mathrm{vec}(\mathbf{U}_r\bm{\Gamma}\mathbf{V}_r^{\mathsf H}),
\end{equation}
where the real-valued parameter vector is
\begin{equation*}
\begin{aligned}
\bm{\zeta}
=
\big[
&\mathbf{d}^{\mathsf T},
\bm{\phi}^{\mathsf T},
\Re\{\mathbf{a}_1\}^{\mathsf T},
\Im\{\mathbf{a}_1\}^{\mathsf T},
\Re\{\mathbf{s}_1\}^{\mathsf T},
\Im\{\mathbf{s}_1\}^{\mathsf T},
\bm{\gamma}^{\mathsf T},\\
&\Re\{\mathrm{vec}(\mathbf{U}_r)\}^{\mathsf T},
\Im\{\mathrm{vec}(\mathbf{U}_r)\}^{\mathsf T},\\
&\Re\{\mathrm{vec}(\mathbf{V}_r)\}^{\mathsf T},
\Im\{\mathrm{vec}(\mathbf{V}_r)\}^{\mathsf T}
\big]^{\mathsf T}\in\mathbb{R}^{n_{\zeta}}.
\end{aligned}
\end{equation*}
with $n_{\zeta}=4P_1+2L+r+2Nr+2Lr$. The derivatives in $\mathbf{F}(\bm{\zeta})\in\mathbb{R}^{n_{\zeta}\times n_{\zeta}}$ are taken with respect to the entries of $\bm{\zeta}$. The set of constraints can be represented as a function $\mathbf{g}(\bm{\zeta})=\mathbf{0}$ with $\mathbf{g}\in\mathbb{R}^{3r+2}$. We define its Jacobian as $\mathbf{G}=\partial\mathbf{g}(\bm{\zeta})/\partial\bm{\zeta}^{\mathsf T}\in\mathbb{R}^{(3r+2)\times n_{\zeta}}$. If $\mathbf{N}_G\in\mathbb{R}^{n_{\zeta}\times(n_{\zeta}-3r-2)}$ is an orthonormal basis for the nullspace of $\mathbf{G}$, then the constrained bound is
\begin{equation}
\mathbf{PR\text{-}CRB}_u^{(r)}(\bm{\zeta})
=
\mathbf{N}_G
\left(
\mathbf{N}_G^{\mathsf T}\mathbf{F}(\bm{\zeta})\mathbf{N}_G
\right)^{-1}
\mathbf{N}_G^{\mathsf T}.
\end{equation}
The corresponding channel-domain bound is obtained as in the known-pilot case.


\section{Numerical Results}

This section assesses the proposed \ac{PR}-based \ac{NF} multipath channel estimation framework under two representative propagation regimes. We first consider a pure NLoS setting, where no dominant LoS component is present, and evaluate the impact of \ac{SNR} and the number of users on the \ac{NMSE} performance. We then consider a mixed LoS/NLoS setting, where the effects of \ac{SNR} and LoS dominance are investigated through the Rician factor.

\paragraph*{Channel and power model}

We consider the \ac{NF} multipath uplink model introduced in Section~\ref{sec:system_model}, where all $K$ users transmit simultaneously to the BS. The gain of the $p$-th path of user $k$ is modeled as $\alpha_k^{(p)}=\sqrt{P_k^{(p)}}e^{j\psi_k^{(p)}}$, where $P_k^{(p)}$ and $\psi_k^{(p)}$ denote the corresponding path power and phase, respectively. For a fair comparison across users and channel realizations, the path powers of each user are normalized such that $\sum_{p=1}^{P_k} P_k^{(p)} = P_{\mathrm{tot}}$, with $P_{\mathrm{tot}}=1$. The nominal \ac{SNR} is defined as $P_s/\sigma^2$, where $P_s$ is the transmit power and $\sigma^2$ is the noise variance. In the multi-user setting, increasing $K$ also increases the aggregate contribution of the interfering users, which is treated by the proposed \ac{PR} estimators as a relaxed nuisance component. Therefore, the user-scaling results reflect both the effect of multi-user interference and the ability of the proposed estimators to separate the desired-user multipath channel from the remaining users. In the simulations, channels are generated with $P_{\max}$ paths per user, whereas the number of estimated paths is determined adaptively by each algorithm. For pure NLoS channels, the path powers are randomly generated and then normalized. For Rician LoS/NLoS channels, the power allocation is controlled by $K_R=P_{\mathrm{LoS}}/P_{\mathrm{NLoS}}$, with $P_{\mathrm{LoS}}+P_{\mathrm{NLoS}}=P_{\mathrm{tot}}$.

\paragraph*{Simulation setup}

Azimuth angles are drawn uniformly from $[0^\circ,120^\circ]$, while ranges are drawn from the radiative \ac{NF} region, i.e., $[3.70,68.45]$ m. The path-search-based estimators use an adaptive stopping threshold $\varepsilon$ and a maximum number of extracted paths $P_{\max}=10$, with $\varepsilon=10^{-2}$ unless otherwise stated. Performance is evaluated in terms of the NMSE between the true and estimated channels. Comparisons also include the classical parametric \ac{CRB} in \cite{gurgunoglu2025_2dmusic}. The remaining simulation parameters are summarized in Table~\ref{tab:sim_params}.

\begin{table}[t]
\caption{Simulation Parameters}
\label{tab:sim_params}
\centering
\footnotesize
\renewcommand{\arraystretch}{1.1}
\begin{tabular}{p{0.35\columnwidth}cccc}
\hline
\textbf{Parameter (Symbol)} & \textbf{Fig.~2} & \textbf{Fig.~3} & \textbf{Fig.~4} & \textbf{Fig.~5} \\
\hline
Number of antennas ($N$) & \multicolumn{4}{c}{$38$} \\
Carrier frequency ($f_{\mathrm{c}}$) & \multicolumn{4}{c}{$3\,\mathrm{GHz}$} \\
\hline
Number of users ($K$) & $2$ & $1$--$4$ & $2$ & $2$ \\
Number of snapshots ($L$) & $80$ & $80$ & $40$ & $40$ \\
Rician factor ($K_R$) (dB) & $0$ & $0$ & $10$ & $0$--$15$ \\
SNR (dB) & $[-10,25]$ & $18$ & $[-10,25]$ & $18$ \\
\hline
\end{tabular}
\end{table}

\paragraph*{Results}

Figs \ref{fig:snr_nlos} and \ref{fig:nlos_users} first examine the pure NLoS case ($K_R=0$), where the channel is composed only of scattered multipath components without a dominant LoS path. Fig \ref{fig:snr_nlos} shows that the proposed \ac{PR}-based estimators achieve a consistent NMSE reduction as the \ac{SNR} increases and remain close to the corresponding \acp{CRB}. In contrast, \ac{2D-MUSIC} exhibits a noticeably larger performance gap, especially at moderate and high \ac{SNR}. Fig.\ref{fig:nlos_users} further evaluates the robustness of the estimators under increasing multi-user interference. Although the NMSE of all methods degrades as the number of users increases, the proposed \ac{PR} estimators maintain the lowest NMSE across the considered values of $K$. This confirms that the proposed estimators can effectively relax the aggregate nuisance component while preserving accurate estimation of the desired-user multipath channel.

Figs \ref{fig:snr_los} and \ref{fig:kfactor} then consider the mixed LoS/NLoS case, where the channel contains both a dominant LoS component and scattered NLoS paths. Fig.\ref{fig:snr_los} shows that the presence of a dominant LoS component ($K_R=10$) improves the estimation accuracy and brings the proposed methods closer to the corresponding bounds over the considered \ac{SNR} range. Fig.\ref{fig:kfactor} further confirms this behavior by showing that increasing the Rician factor leads to a clear NMSE improvement, since the channel becomes more dominated by the structured LoS component.
Overall, the results show that the proposed \ac{PR}-based estimators consistently outperform \ac{2D-MUSIC} across all considered scenarios. As expected, \ac{PR-ML} achieves the best performance because it exploits the known pilot sequence. More importantly, \ac{PR-CF} also provides a clear performance advantage over \ac{2D-MUSIC} under the same unknown-symbol setting, demonstrating that the proposed covariance-fitting formulation can achieve accurate \ac{NF} multipath channel estimation even without prior knowledge of the transmitted symbols.

\begin{figure}[t]
\centering
%
\definecolor{mycolor1}{rgb}{0.06600,0.44300,0.74500}%
\definecolor{mycolor2}{rgb}{0.86600,0.32900,0.00000}%
\definecolor{mycolor3}{rgb}{1.00000,0.00000,1.00000}%
\definecolor{mycolor4}{rgb}{0.00000,1.00000,1.00000}%
\definecolor{mycolor5}{rgb}{0.12941,0.12941,0.12941}%
\begin{tikzpicture}

\begin{axis}[
width=0.81\linewidth,
height=0.59\columnwidth,
scale only axis,
xmin=-10,
xmax=25,
xlabel={SNR (dB)},
ylabel={NMSE},
ymode=log,
ymin=1e-6,
ymax=10,
yminorticks=true,
xlabel style={font=\small},
ylabel style={font=\small},
axis background/.style={fill=white},
grid=both,
major grid style={draw=gray!5},
minor grid style={draw=gray!25},
xmajorgrids,
ymajorgrids,
yminorgrids,
tick label style={font=\small},
legend style={
at={(0,0)},
anchor=south west,
legend columns=2,
legend cell align=left,
font=\scriptsize,
row sep=1pt,
column sep=1pt
}
]
\addplot [color=mycolor1, line width=1.6pt, mark=triangle, mark options={solid, mycolor1}]
  table[row sep=crcr]{%
-10	9.49410774059649\\
-5	2.73609980177842\\
0	0.904501096220228\\
5	0.30497797331883\\
10	0.197458215729124\\
15	0.1508181312929732\\
20	0.0853783241981285\\
25	0.0460478483193521\\
};
\addlegendentry{2D-MUSIC}

\addplot [color=mycolor2, line width=1.6pt, mark=o, mark options={solid, mycolor2}]
  table[row sep=crcr]{%
-10	0.110066348288018\\
-5	0.0303602923409322\\
0	0.010174869981856\\
5	0.00323995769062225\\
10	0.00120381745026523\\
15	0.000428716134962259\\
20	0.000203004180024203\\
25	0.000102559990137966\\
};
\addlegendentry{PR-CF}

\addplot [color=black, line width=1.6pt]
  table[row sep=crcr]{%
-10	0.063247096622237\\
-5	0.0163545306652148\\
0	0.00489552801011054\\
5	0.00158857866287782\\
10	0.000573565190932491\\
15	0.000162767337315074\\
20	5.98803678844424e-05\\
25	1.63597366673247e-05\\
};
\addlegendentry{PR-ML}

\addplot [color=mycolor3, dashed, line width=1.6pt, mark size=2.5pt, mark=square, mark options={solid, mycolor3}]
  table[row sep=crcr]{%
-10	0.0289451100402794\\
-5	0.00915324748514871\\
0	0.00289451100402775\\
5	0.000915324748514437\\
10	0.000289451100402895\\
15	9.15324748514557e-05\\
20	2.89451100402699e-05\\
25	9.15324748514991e-06\\
};
\addlegendentry{$\text{PR-CRB}_\text{P}$}

\addplot [color=mycolor4, dashdotted, line width=1.6pt]
  table[row sep=crcr]{%
-10	0.0349813697167899\\
25	1.1062080395822e-05\\
};
\addlegendentry{$\text{PR-CRB}_\text{U}$}

\addplot [color=green, dashed, line width=1.6pt]
  table[row sep=crcr]{%
-10	0.0202615770281956\\
25	6.40727323960494e-06\\
};
\addlegendentry{Parametric CRB}

\end{axis}
\end{tikzpicture}%
\caption{Pure NLoS (\(K_R=0\)): effect of SNR for \(K=2\) and \(L=80\).}
\label{fig:snr_nlos}
\end{figure}
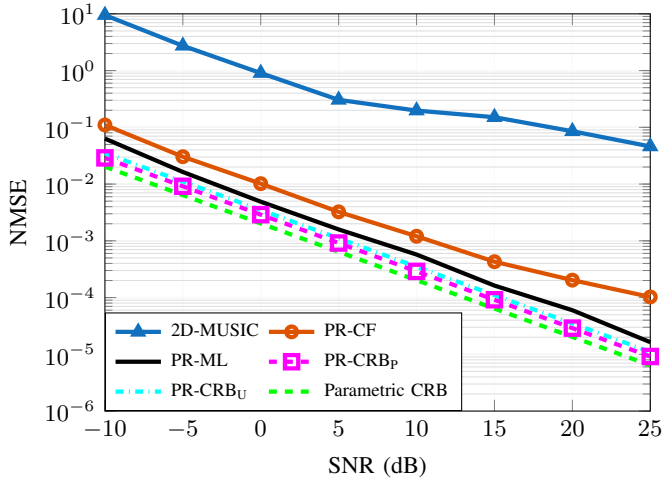

\begin{figure}[t]
\centering
%
\definecolor{mycolor1}{rgb}{0.00000,0.44700,0.74100}%
\definecolor{mycolor2}{rgb}{0.85000,0.32500,0.09800}%
\definecolor{mycolor3}{rgb}{1.00000,0.00000,1.00000}%
\definecolor{mycolor4}{rgb}{0.00000,1.00000,1.00000}%
\definecolor{mycolor5}{rgb}{0.12941,0.12941,0.12941}%
\begin{tikzpicture}

\begin{axis}[
width=0.81\columnwidth,
height=0.59\columnwidth,
at={(0\columnwidth,0\columnwidth)},
scale only axis,
bar shift auto,
log origin=infty,
xmin=0.5,
xmax=4.5,
xtick={1,2,3,4},
xlabel style={font=\color{mycolor5}},
xlabel={Number of Users \(K\)},
ymode=log,
ymin=1e-06,
ymax=10,
yminorticks=true,
ylabel style={font=\color{mycolor5}},
ylabel={NMSE},
axis background/.style={fill=white},
major grid style={draw=gray!5},
minor grid style={draw=gray!25},
xmajorgrids,
ymajorgrids,
yminorgrids,
legend style={
font=\scriptsize,
at={(0.01,1)},
anchor=north west,
legend columns=2,
legend cell align=left,
row sep=1pt,
column sep=4pt,
draw=black,
fill=white
}
]
\addplot[ybar, bar width=0.12, fill=mycolor1, area legend] table[row sep=crcr] {%
1	0.0523739435616713\\
2	0.0772503778510217\\
3	0.134892050489917\\
4	0.379397953551836\\
};
\addlegendentry{2D-MUSIC}

\addplot[ybar, bar width=0.12, fill=mycolor2, area legend] table[row sep=crcr] {%
1	0.000200972489559112\\
2	0.00037726658848836\\
3	0.000621470966382801\\
4	0.00154321242037363\\
};
\addlegendentry{PR-CF}

\addplot[ybar, bar width=0.12, fill=black, area legend] table[row sep=crcr] {%
1	0.000118389183000088\\
2	0.000141036850902498\\
3	0.000206699652128366\\
4	0.000398004053991786\\
};
\addlegendentry{PR-ML}

\addplot[ybar, bar width=0.12, fill=mycolor3, area legend] table[row sep=crcr] {%
1	6.54161379431694e-05\\
2	8.74117279646957e-05\\
3	9.85670332569474e-05\\
4	0.000141826697956912\\
};
\addlegendentry{$\text{PR-CRB}_\text{P}$}

\addplot[ybar, bar width=0.12, fill=mycolor4, area legend] table[row sep=crcr] {%
1	5.61962582425301e-05\\
2	7.31498960863538e-05\\
3	8.31847866320421e-05\\
4	0.000120831289824929\\
};
\addlegendentry{$\text{PR-CRB}_\text{U}$}

\addplot[ybar, bar width=0.12, fill=black!50!green, area legend] table[row sep=crcr] {%
1	3.93373807697711e-05\\
2	5.12049272604477e-05\\
3	5.82293506424295e-05\\
4	8.45819028774504e-05\\
};
\addlegendentry{Parametric CRB}

\end{axis}
\end{tikzpicture}%
\caption{Pure NLoS (\(K_R=0\)): effect of \(K\) at SNR \(=18\) dB, \(L=80\).}
\label{fig:nlos_users}
\end{figure}
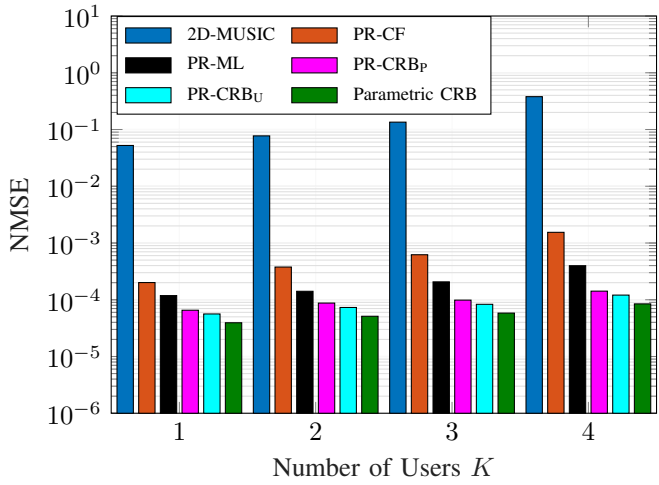

\begin{figure}[t]
\centering
%
\definecolor{mycolor1}{rgb}{0.06600,0.44300,0.74500}%
\definecolor{mycolor2}{rgb}{0.86600,0.32900,0.00000}%
\definecolor{mycolor3}{rgb}{1.00000,0.00000,1.00000}%
\definecolor{mycolor4}{rgb}{0.00000,1.00000,1.00000}%
\definecolor{mycolor5}{rgb}{0.12941,0.12941,0.12941}%
\begin{tikzpicture}

\begin{axis}[
width=0.81\columnwidth,
height=0.59\columnwidth,
scale only axis,
xmin=-10,
xmax=25,
xlabel={SNR (dB)},
ylabel={NMSE},
ymode=log,
ymin=1e-6,
ymax=10,
yminorticks=true,
xlabel style={font=\small},
ylabel style={font=\small},
axis background/.style={fill=white},
grid=both,
major grid style={draw=gray!5},
minor grid style={draw=gray!25},
xmajorgrids,
ymajorgrids,
yminorgrids,
tick label style={font=\small},
legend style={
at={(0,0)},
anchor=south west,
legend columns=2,
legend cell align=left,
font=\scriptsize,
row sep=0.5pt,
column sep=0.5pt
}
]
\addplot [color=mycolor1, line width=1.8pt, mark=triangle, mark options={solid, mycolor1}]
  table[row sep=crcr]{%
-10	1.34474681537573\\
-5	0.417959166356877\\
0	0.123126077624958\\
5	0.0391776897885193\\
10	0.0111667146701579\\
15	0.00385396357973915\\
20	0.00160919724686633\\
25	0.000634357077786357\\
};
\addlegendentry{2D-MUSIC}

\addplot [color=mycolor2, line width=1.8pt, mark=o, mark options={solid, mycolor2}]
  table[row sep=crcr]{%
-10	0.122476157958641\\
-5	0.0257260146140955\\
0	0.00409030544587462\\
5	0.00105141940701797\\
10	0.000151460791985311\\
15	5.07091165605781e-05\\
20	1.76999595996955e-05\\
25	6.06009928864337e-06\\
};
\addlegendentry{PR-CF}

\addplot [color=black, line width=2.2pt]
  table[row sep=crcr]{%
-10	0.0138933695047799\\
-5	0.00458260115292451\\
0	0.00130611857713464\\
5	0.00044796498913617\\
10	0.000129602841891819\\
15	4.23979384656671e-05\\
20	1.32054870238983e-05\\
25	4.62734643263777e-06\\
};
\addlegendentry{PR-ML}

\addplot [color=mycolor3, dashed, line width=2.6pt, mark size=2.5pt, mark=square, mark options={solid, mycolor3}]
  table[row sep=crcr]{%
-10	0.0134510880728228\\
-5	0.00425360753177262\\
0	0.0013451088072891\\
5	0.000425360753141871\\
10	0.000134510880718779\\
15	4.253607531799e-05\\
20	1.34510880736401e-05\\
25	4.25360753201285e-06\\
};
\addlegendentry{$\text{PR-CRB}_\text{P}$}

\addplot [color=mycolor4, dashdotted, line width=2.0pt]
  table[row sep=crcr]{%
-10	0.0154522238215021\\
25	4.88642221929558e-06\\
};
\addlegendentry{$\text{PR-CRB}_\text{U}$}

\addplot [color=green, dashed, line width=2.0pt]
  table[row sep=crcr]{%
-10	0.00941576165097594\\
25	2.977525272409e-06\\
};
\addlegendentry{Parametric CRB}

\end{axis}
\end{tikzpicture}%
\caption{LoS/NLoS: effect of SNR for \(K_R=10\) and \(L=40\).}
\label{fig:snr_los}
\end{figure}
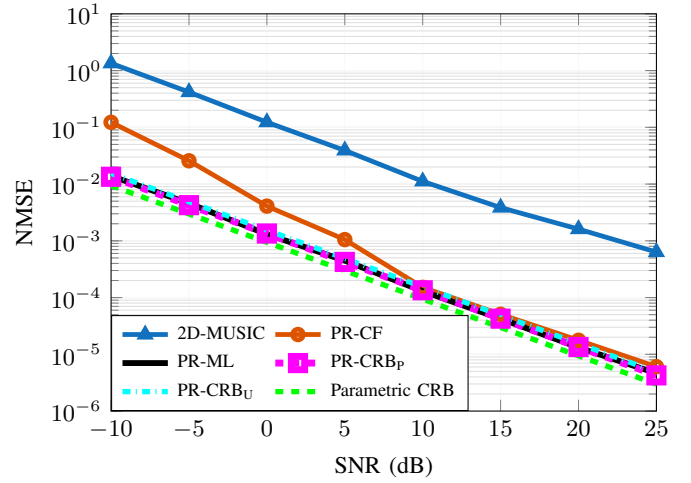

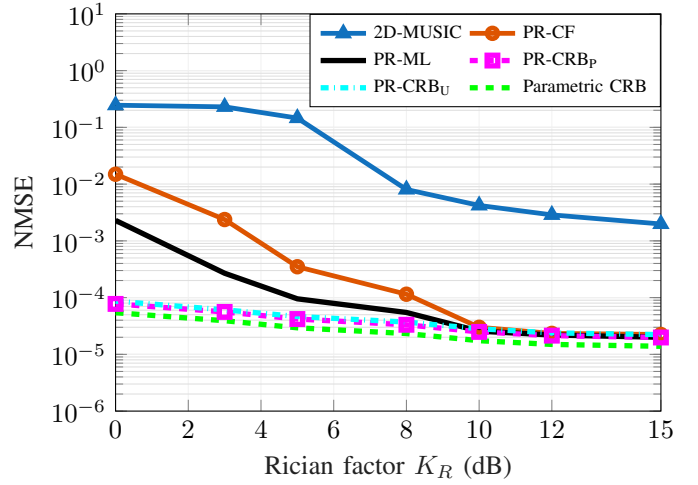
\begin{figure}[t]
    \centering
%
\definecolor{mycolor1}{rgb}{0.06600,0.44300,0.74500}%
\definecolor{mycolor2}{rgb}{0.86600,0.32900,0.00000}%
\definecolor{mycolor3}{rgb}{1.00000,0.00000,1.00000}%
\definecolor{mycolor4}{rgb}{0.00000,1.00000,1.00000}%
\definecolor{mycolor5}{rgb}{0.12941,0.12941,0.12941}%
\begin{tikzpicture}

\begin{axis}[%
width=0.81\columnwidth,
height=0.59\columnwidth,
at={(0\columnwidth,0\columnwidth)},
scale only axis,
xmin=0,
xmax=15,
xtick={0,2,4,6,8,10,12,15},
xlabel style={font=\color{mycolor5}},
xlabel={Rician factor $K_R$ (dB)},
ymode=log,
ymin=1e-06,
ymax=10,
yminorticks=true,
ylabel style={font=\color{mycolor5}},
ylabel={NMSE},
axis background/.style={fill=white},
title style={font=\bfseries\color{mycolor5}},
major grid style={draw=gray!10},
minor grid style={draw=gray!25},
xmajorgrids,
ymajorgrids,
yminorgrids,
legend style={
at={(1,1)},
anchor=north east,
legend columns=2,
legend cell align=left,
align=left,
font=\scriptsize
}
]
\addplot [color=mycolor1, line width=1.8pt, mark=triangle, mark options={solid, mycolor1}]
  table[row sep=crcr]{%
0	0.24574084591293\\
3	0.229891786732846\\
5	0.145773841241077\\
8	0.00803953998270023\\
10	0.00421119998964204\\
12	0.00287001009021635\\
15	0.00198434860495384\\
};
\addlegendentry{2D-MUSIC}

\addplot [color=mycolor2, line width=1.8pt, mark=o, mark options={solid, mycolor2}]
  table[row sep=crcr]{%
0	0.0149352652790928\\
3	0.00237587337423947\\
5	0.000350019964470567\\
8	0.000114737378324117\\
10	2.9986195635214e-05\\
12	2.34225132773745e-05\\
15	2.24095431566372e-05\\
};
\addlegendentry{PR-CF}

\addplot [color=black, line width=2.0pt]
  table[row sep=crcr]{%
0	0.00229441419890557\\
3	0.00026878694233397\\
5	9.52004476648116e-05\\
8	5.4722817673842e-05\\
10	2.56266142830381e-05\\
12	2.1935692913896e-05\\
15	1.99956916181727e-05\\
};
\addlegendentry{PR-ML}

\addplot [color=mycolor3, dashed, line width=2.0pt, mark=square, mark options={solid, mycolor3}]
  table[row sep=crcr]{%
0	7.74728108786002e-05\\
3	5.59364864333473e-05\\
5	4.19412702014832e-05\\
8	3.32231604889664e-05\\
10	2.48720217640754e-05\\
12	2.13817579192748e-05\\
15	1.98233589356836e-05\\
};
\addlegendentry{$\text{PR-CRB}_\text{P}$}

\addplot [color=mycolor4, dashdotted, line width=2.0pt]
  table[row sep=crcr]{%
0	8.66243161705223e-05\\
3	6.08702988982797e-05\\
5	4.64994010971019e-05\\
8	3.70698949807651e-05\\
10	2.82990030673265e-05\\
12	2.39022539690955e-05\\
15	2.22627221525832e-05\\
};
\addlegendentry{$\text{PR-CRB}_\text{U}$}

\addplot [color=green, dashed, line width=2.0pt]
  table[row sep=crcr]{%
0	5.42309676150202e-05\\
3	3.91555405033431e-05\\
5	2.93588891410382e-05\\
8	2.32562123422765e-05\\
10	1.74104152348528e-05\\
12	1.49672305434924e-05\\
15	1.38763512549785e-05\\
};
\addlegendentry{Parametric CRB}

\end{axis}
\end{tikzpicture}%
    \caption{LoS/NLoS: effect of \(K_R\) at SNR \(=18\) dB and \(L=40\).}
    \label{fig:kfactor}
\end{figure}


\section{Conclusions}

In this paper, we studied \ac{NF} NLoS multipath channel estimation for multi-user uplink systems with large-aperture arrays. Within a \ac{PR} framework, the desired-user channel was represented by a structured multipath model, while the remaining users were absorbed into a relaxed nuisance component. For known pilots, we developed a \ac{PR-ML} estimator and derived the corresponding \ac{PRCRBp}. For unknown symbols, we proposed a rank-adaptive \ac{PR-CF} formulation and derived the associated \ac{PRCRBu}. Numerical results showed that the resulting estimators remain close to their bounds and consistently outperform near-field \ac{2D-MUSIC} in both pure NLoS and mixed LoS/NLoS scenarios. In particular, \ac{PR-CF} achieved a clear advantage in the unknown-symbol setting, highlighting its suitability for practical \ac{NF} channel estimation when pilot or symbol knowledge is unavailable. These results indicate that the proposed framework can serve as an effective physical-layer solution for future large-aperture uplink systems.

\bibliographystyle{IEEEtran}
\bibliography{refs}

\end{document}